\documentclass[conference]{IEEEtran}
\IEEEoverridecommandlockouts

\makeatletter
\def\bstctlcite{\@ifnextchar[{\@bstctlcite}{\@bstctlcite[@auxout]}}
\def\@bstctlcite[#1]#2{\@bsphack
  \@for\@citeb:=#2\do{%
    \edef\@citeb{\expandafter\@firstofone\@citeb}%
    \if@filesw\immediate\write\csname #1\endcsname{\string\citation{\@citeb}}\fi}%
  \@esphack}
\makeatother

\usepackage{amsmath}
\usepackage{amssymb}
\usepackage{amsfonts}
\usepackage{algorithm}
\usepackage{algpseudocode}
\usepackage{graphicx}
\usepackage{textcomp}
\usepackage{multirow}
\usepackage{xcolor}
\usepackage{cite}
\usepackage{tabularx}
\usepackage{booktabs}
\usepackage{lipsum}
\usepackage{url}
\usepackage{soul}
\usepackage{stfloats}
\usepackage{etoolbox}

\usepackage{mathtools}

\def\BibTeX{{\rm B\kern-.05em{\sc i\kern-.025em b}\kern-.08emT\kern-.1667em\lower.7ex\hbox{E}\kern-.125emX}}

\begin{document}
\bstctlcite{IEEEexample:BSTcontrol}

\title{
FPGA Resource-aware Structured Pruning\\for Real-Time Neural Networks
\thanks{This work is supported by Engineering and Physical Sciences Research Council (EPSRC) grant EP/S030069/1 and NSF Institute A3D3, NSF 2117997.}
}

\author{
\IEEEauthorblockN{Benjamin Ramhorst\IEEEauthorrefmark{1}}
\IEEEauthorblockA{\textit{Electrical and Electronic Engineering} \\
\textit{Imperial College London}\\
London, United Kingdom \\
benjamin.ramhorst@inf.ethz.ch}

\and

\IEEEauthorblockN{Vladimir Lončar\IEEEauthorrefmark{2}}
\IEEEauthorblockA{\textit{Laboratory for Nuclear Science} \\
\textit{Massachusetts Institute of Technology} \\
Cambridge, United States of America \\
vloncar@mit.edu}
\thanks{\IEEEauthorrefmark{1} Currently at ETH Zürich, Switzerland}
\thanks{\IEEEauthorrefmark{2} Also at Institute of Physics Belgrade, Serbia}

\and 

\IEEEauthorblockN{George A. Constantinides}
\IEEEauthorblockA{\textit{Electrical and Electronic Engineering} \\
\textit{Imperial College London}\\
London, United Kingdom \\
g.constantinides@imperial.ac.uk}

}
\maketitle

\begin{abstract}
Neural networks achieve state-of-the-art performance in image classification, speech recognition, scientific analysis and many more application areas. Due to the high computational complexity and memory footprint of neural networks, various compression techniques, such as pruning and quantization, have been proposed in literature. Pruning sparsifies a neural network, reducing the number of multiplications and memory. However, pruning often fails to capture properties of the underlying hardware, causing unstructured sparsity and load-balance inefficiency, thus bottlenecking resource improvements. 
We propose a hardware-centric formulation of pruning, by formulating it as a knapsack problem with resource-aware tensor structures. Evaluated on a range of tasks, including sub-microsecond particle classification at CERN's Large Hadron Collider and fast image classification, the proposed method achieves reductions ranging between 55\% and 92\% in the DSP utilization and up to 81\% in BRAM utilization.
\end{abstract}


\section{Introduction}
Modern neural networks are associated with high computational complexity and memory footprint, linked to the high dimensionality of weight tensors. With the ever-increasing need for faster computation and lower power consumption, driven by real-time systems and Internet-of-Things (IoT), neural networks must be adapted to reduce their complexity. A common method for network compression is pruning, the process of sparsifying a neural network by setting some of its weights to zero. In their seminal work, Han~\emph{et al.}~\cite{Han_Pruning} applied iterative, magnitude-based unstructured pruning to modern neural networks. The networks were trained with regularization, shifting weights towards zero. Weights below a certain magnitude were pruned, while the remaining weights were updated, thus capturing the loss of the pruned weights. While unstructured pruning achieves high sparsities, little improvements in resource utilization and inference latency are observed~\cite{Hoefler_PruningSurvey}, due to the overhead of sparse encodings.

\begin{figure}
\centering
\includegraphics[width=0.35\textwidth]{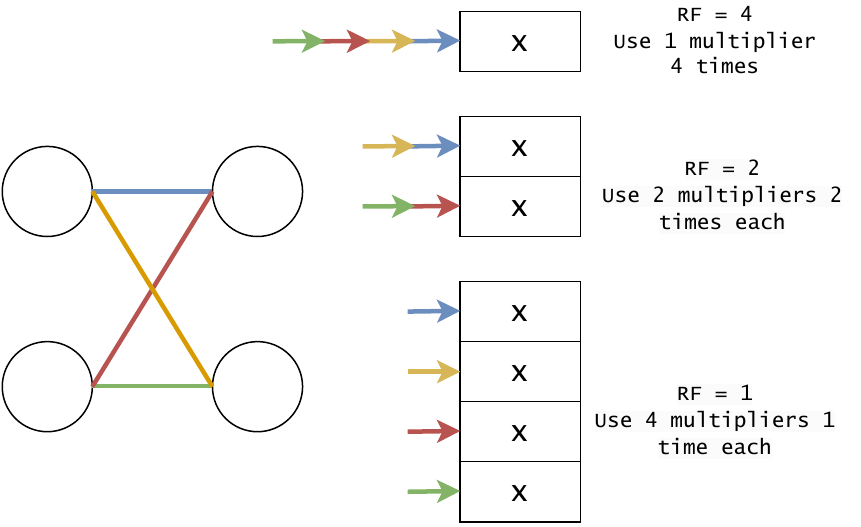}
\caption{Variations of RF and the impact on resource utilization~\cite{Duarte_hls4ml}.}
\label{fig:hls4ml_rf}
\end{figure}

\texttt{hls4ml}~\cite{GitHub_hls4ml, Duarte_hls4ml} is an open-source library for real-time inference of neural networks on FPGAs. Originally intended for sub-microsecond data filtering in high-energy physics, \texttt{hls4ml} has become a versatile tool for real-time inference, with a full on-chip design and high parallelism, frequently leading to high resource utilization. A key variable in \texttt{hls4ml} is the \emph{Reuse Factor} (RF), determining the trade-off between latency and resource utilization, as shown in Fig.~\ref{fig:hls4ml_rf}. In \texttt{hls4ml}, matrix-vector multiplication is the baseline for fully connected layers, as well as convolutional layers, implemented using the \emph{im2col} transformation.

We propose an FPGA resource-aware structured pruning algorithm, able to capture, at training time, the underlying mapping of network weights to digital signal processing blocks (DSPs) and block random access memory (BRAM). Groups of weights are iteratively removed through constrained optimization, formulated as a knapsack problem. The methodology described in this research has been integrated with \texttt{hls4ml} and open-sourced with an Apache 2.0 license\footnote{GitHub page: https://github.com/fastmachinelearning/hls4ml/tree/hardware-aware-pruning}.

\section{Resource-aware pruning}
\label{sec:method}

To maximize resource savings of accelerated neural networks, we consider the optimization of DSP and BRAM utilization, by pruning all the weights processed by the same multiplier or block of memory. We refer to these groups of weights as resource-aware vectors of weights. The mapping between weights and DSP blocks is deterministic from the RF. Given a weight matrix, weights processed by the same multiplier can be obtained by transposing and flattening the matrix into a vector. The vector is split into sub-vectors of length RF, with each sub-vector representing the weights processed by the same DSP block. Weights processed by the same DSP are stored as subsequent words in the same block of memory, typically implemented as 1K x 36. Weights are quantized before deployment to a lower precision, so each block of RAM stores weights processed by several neighboring DSP blocks. As an example, groups of weights quantized to 18 bits would map to 2 DSP blocks for each block of RAM.



\texttt{hls4ml} supports heterogeneous configuration of layers, allowing fine-grained tuning of the hardware-configuration, by varying the RF and precision. Furthermore, Vivado aims to implement multiplications for precisions lower than 10 bits through LUTs, rather than DSPs. As a result of these factors, pruning becomes a non-trivial optimization problem, aimed at minimizing network loss and overall resource consumption. We solve this problem by transforming it to a knapsack problem, similar to the work by Shen \emph{et al.}~\cite{Shen_HALP} on latency-aware pruning for GPUs. Given a a set of resource-aware vectors $\mathbb{W} = \{\mathbf{w}_1, \hdots, \mathbf{w}_n\}$ and the vector-valued function $R: \mathbb{R}^k \rightarrow \mathbb{R}^m$, representing the resource consumption (DSP, BRAM) of a group of weights, we formulate pruning as an optimization problem (Eq. \ref{eq:pruning_general_formula}) of selecting a subset of weights, $\hat{\mathbb{W}} \subseteq
\mathbb{W}$, minimizing network loss while maintaining a resource budget, $\mathbf{c}$. To minimize network loss, we aim to maximize the normalized\footnote{Vector norms are normalized in every layer, avoiding bias to highly-dimensional layers.} magnitude of the selected weights, similar to magnitude-based pruning~\cite{Han_Pruning}.
\begin{gather}
    \max_{\hat{\mathbb{W}}} \; \sum_{\mathbf{w}_i \in \hat{\mathbb{W}}} \frac{\lVert \mathbf{w}_i \rVert_1}{\max_{\mathbf{w}_i, \mathbf{w}_j \in L}  {\lVert \mathbf{w}_j \rVert}_1} \nonumber \\
    \text{subject to} \sum_{\mathbf{w}_i \in \hat{\mathbb{W}}} R(\mathbf{w}_i) \preceq \mathbf{c} \label{eq:pruning_general_formula} \\  
    \hat{\mathbb{W}} \subseteq \mathbb{W} \nonumber
\end{gather}
To solve the combinatorial optimization problem in Equation~\ref{eq:pruning_general_formula}, we consider the 0-1 knapsack problem:
\begin{gather}
    \max_{\boldsymbol{x}} \boldsymbol{v}^T \boldsymbol{x} \nonumber \\ 
    \text{subject to} \; \boldsymbol{U} \boldsymbol{x} \preceq \boldsymbol{c}  \label{eq:knapsack_nd} \\
    \boldsymbol{x} \in \{0, 1\}^n \nonumber 
\end{gather}
In our case, the vector $\boldsymbol{v}$ corresponds to the normalized magnitudes, while the matrix $\boldsymbol{U}$ corresponds to the resource utilization. We find that the knapsack problem can be efficiently solved using branch-and-cut methods and the solution is such that:
\begin{gather}
    x_i = 
    \begin{cases}
        0 & \text{if }\mathbf{w}_i\text{ pruned} \\
        1 & \text{otherwise}
    \end{cases}
\end{gather}

During training we include a resource-aware regularization loss,  shifting weights sharing the same hardware resource towards zero. Pruning is implemented iteratively, such that the resource budget, $\mathbf{c}$, is repeatedly reduced, until the target resource consumption is achieved, or, network performance on the validation set drops below the relative tolerance. At each iteration, the remaining weights are retrained.

\section{Results}
\label{sec:results}
We evaluate the effects of resource-aware pruning on three  benchmark models: high-energy jet tagging \cite{Duarte_hls4ml}, SVHN classification, as described by Aarrestad \emph{et al.}~\cite{Aarrestad_CNN} and Fashion MNIST classification \cite{fashion_mnist_data} with a LeNet-like \cite{LeNet} architecture. The models are trained using \texttt{Keras}~\cite{Keras} with a \texttt{TensorFlow} backend~\cite{TensorFlow}, by minimizing categorical cross-entropy and synthesized using Vivado, Vivado HLS 2020.1 and \texttt{hls4ml} 0.7.1. The resource utilization is reported post place and route with the target device set to Virtex UltraScale+ XCVU9P. The results are reported in Table~\ref{tab:results_summary}. For particle and SVHN classification, we vary the RF, and report a range for DSP and BRAM reduction. Due to the size of LeNet, we manually tune the RF of every layer in Fashion MNIST classification, in order to meet Vivado partition and unroll limits. We also report the difference in accuracy between the pruned model with the lowest accuracy and the baseline model.

\begin{table}
\caption{Effects of resource-aware pruning.}
\centering
\begin{tabular}{|c|c|c|c|}
\hline
Model & Jets & SVHN & Fashion MNIST \\ \hline
Baseline accuracy [\%] & 76.61 & 91.95 & 89.72 \\ \hline
RF & 2, 4, 8, 16 & 3, 9, 27 & Fixed per layer \\ \hline
DSP reduction & 5.8x - 12.2x & 2.2x - 3.9x & 4.7x  \\ \hline
BRAM reduction & 2.3x - 5.2x & 1.4x & 2.1x  \\ \hline
$\Delta$ accuracy [pp] & $-$0.63 & $+$0.26 & $+$0.02  \\ \hline
\end{tabular}
\label{tab:results_summary}
\end{table}


\bibliographystyle{IEEEtran}
\bibliography{main}

\end{document}